\documentclass[twoside,twocolumn,9pt]{article}
\usepackage{extsizes}
\usepackage[super,sort&compress,comma]{natbib} 
\usepackage[version=3]{mhchem}
\usepackage[left=1.5cm, right=1.5cm, top=1.785cm, bottom=2.0cm]{geometry}
\usepackage{balance}
\usepackage{amsmath}
\usepackage{amsfonts}
\usepackage{amssymb}
\usepackage{amssymb}
\usepackage{sectsty}
\usepackage{graphicx} 
\usepackage{lastpage}
\usepackage{ulem}
\usepackage[format=plain,justification=justified,singlelinecheck=false,font={stretch=1.125,small,sf},labelfont=bf,labelsep=space]{caption}
\usepackage{float}
\usepackage{fancyhdr}
\usepackage{fnpos}
\usepackage[english]{babel}
\addto{\captionsenglish}{%
  
}
\usepackage{array}
\usepackage{droidsans}
\usepackage{charter}
\usepackage[T1]{fontenc}
\usepackage[usenames,dvipsnames]{xcolor}
\usepackage{setspace}
\usepackage[compact]{titlesec}
\usepackage{hyperref}

\usepackage{epstopdf}

\definecolor{cream}{RGB}{222,217,201}
\begin{document}

\pagestyle{fancy}
\thispagestyle{plain}
\fancypagestyle{plain}{
\renewcommand{\headrulewidth}{0pt}
}

\makeFNbottom
\makeatletter
\renewcommand\LARGE{\@setfontsize\LARGE{15pt}{17}}
\renewcommand\Large{\@setfontsize\Large{12pt}{14}}
\renewcommand\large{\@setfontsize\large{10pt}{12}}
\renewcommand\footnotesize{\@setfontsize\footnotesize{7pt}{10}}
\makeatother

\renewcommand{\thefootnote}{\fnsymbol{footnote}}
\renewcommand\footnoterule{\vspace*{1pt}%
\color{cream}\hrule width 3.5in height 0.4pt \color{black}\vspace*{5pt}} 
\setcounter{secnumdepth}{5}

\makeatletter 
\renewcommand\@biblabel[1]{#1}            
\renewcommand\@makefntext[1]%
{\noindent\makebox[0pt][r]{\@thefnmark\,}#1}
\makeatother 
\renewcommand{\figurename}{\small{Fig.}~}
\sectionfont{\sffamily\Large}
\subsectionfont{\normalsize}
\subsubsectionfont{\bf}
\setstretch{1.125} 
\setlength{\skip\footins}{0.8cm}
\setlength{\footnotesep}{0.25cm}
\setlength{\jot}{10pt}
\titlespacing*{\section}{0pt}{4pt}{4pt}
\titlespacing*{\subsection}{0pt}{15pt}{1pt}

\fancyfoot{}
\fancyfoot[LO,RE]{\vspace{-7.1pt}\includegraphics[height=9pt]{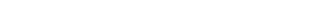}}
\fancyfoot[CO]{\vspace{-7.1pt}\hspace{13.2cm}\includegraphics{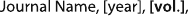}}
\fancyfoot[CE]{\vspace{-7.2pt}\hspace{-14.2cm}\includegraphics{head_foot/RF}}
\fancyfoot[RO]{\footnotesize{\sffamily{1--\pageref{LastPage} ~\textbar  \hspace{2pt}\thepage}}}
\fancyfoot[LE]{\footnotesize{\sffamily{\thepage~\textbar\hspace{3.45cm} 1--\pageref{LastPage}}}}
\fancyhead{}
\renewcommand{\headrulewidth}{0pt} 
\renewcommand{\footrulewidth}{0pt}
\setlength{\arrayrulewidth}{1pt}
\setlength{\columnsep}{6.5mm}
\setlength\bibsep{1pt}

\makeatletter 
\newlength{\figrulesep} 
\setlength{\figrulesep}{0.5\textfloatsep} 

\newcommand{\topfigrule}{\vspace*{-1pt}%
\noindent{\color{cream}\rule[-\figrulesep]{\columnwidth}{1.5pt}} }

\newcommand{\botfigrule}{\vspace*{-2pt}%
\noindent{\color{cream}\rule[\figrulesep]{\columnwidth}{1.5pt}} }

\newcommand{\dblfigrule}{\vspace*{-1pt}%
\noindent{\color{cream}\rule[-\figrulesep]{\textwidth}{1.5pt}} }

\makeatother

\twocolumn[
  \begin{@twocolumnfalse}
{\includegraphics[height=30pt]{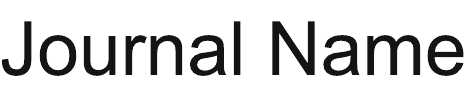}\hfill\raisebox{0pt}[0pt][0pt]{\includegraphics[height=55pt]{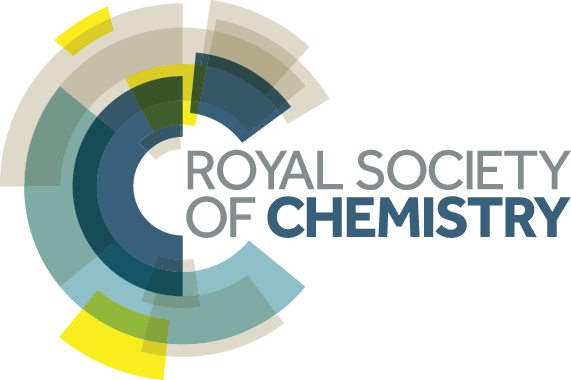}}\\[1ex]
\includegraphics[width=18.5cm]{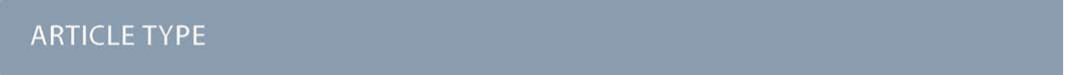}}\par
\vspace{1em}
\sffamily
\begin{tabular}{m{4.5cm} p{13.5cm} }

\includegraphics{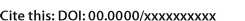} & \noindent\LARGE{\textbf{First principles prediction unveils high-T$_c$ superconductivity in YSc$_2$H$_{24}$ cage structures under pressure}} \\
\vspace{0.3cm} & \vspace{0.3cm} \\

 & \noindent\large{Truong-Tho Pham,\textit{$^{a,b}$} Viet-Ha Chu,\textit{$^c$}} and Duc-Long Nguyen\textit{$^{b,d}$}$^{\ast}$ \\

\includegraphics{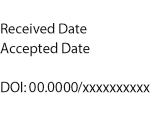} & \noindent\normalsize{The quest for room-temperature superconductivity has been a long-standing aspiration in the field of materials science, driving extensive research efforts. In this work, we present a novel hydride, YSc$_2$H$_{24}$, which is stable at high pressure, identified through crystal structure prediction methods. The discovered material is crystalline in a hexagonal unit cell with space group $P6/mmm$ and has a fastinating structure consisting of two distinct cages: Sc@H$_{24}$ and Y@H$_{30}$. By conducting an extensive numerical investigation of lattice dynamics, electron-phonon coupling, and solving the isotropic Eliashberg equation, we have revealed a significant value of $\lambda$ = 3.27 as the underlying factor responsible for the remarkably high critical temperature (T$_c$) of 302-330 K in YSc$_2$H$_{24}$ at a pressure of 310 GPa. As pressure increases, the T$_c$ remains above the ambient temperature. Our work has the potential to enhance the existing understanding of high-temperature superconductors, with implications for practical applications. The unique network of these cage-like structures holds great promise for advancing our understanding of high-temperature superconductors, potentially leading to innovative applications.} \\
\end{tabular}

 \end{@twocolumnfalse} \vspace{0.6cm}

  ]

\renewcommand*\rmdefault{bch}\normalfont\upshape
\rmfamily
\section*{}
\vspace{-1cm}

\footnotetext{\textit{$^{a}$Laboratory of Magnetism and Magnetic Materials, Science and Technology Advanced Institute, Van Lang University, Ho Chi Minh City, Vietnam}}
\footnotetext{\textit{$^{b}$Faculty of Applied Technology, School of Technology, Van Lang University, Ho Chi Minh City, Vietnam.
\\Email: phamtruongtho@vlu.edu.vn}}
\footnotetext{\textit{$^{c}$Department of Physics, TNU-University of Education, Thai Nguyen, 250000, Vietnam}}
\footnotetext{\textit{$^{d}$Simulation in Materials Science Research Group, Science and Technology Advanced Institute, Van Lang University, Ho Chi Minh City, Vietnam \\Email: nguyenduclong@vlu.edu.vn}}




\section{Introduction}
Considerable focus is devoted to the behaviour of hydrogen under significant pressure in the pursuit of high-temperature superconductivity. In such extreme conditions, the lightest element, hydrogen, is supposed to be capable of forming metallic hydrogen. It has significant high-Tc superconductivity originating from its high Debye temperature and strong electron-phonon interaction \cite{ashcroft1968metallic}. Researchers have investigated the capabilities of hydrogen-rich materials as well as high-pressure solid hydrogen. If one considers these substances to be "chemically precompressed" they may be able to superconduct at considerably lower pressures. It is possible to find superconductivity in compounds that contain a high amount of hydrogen, which also makes these compounds simpler to work with in experiment \cite{semenok2020distribution}.
   The discovery of hydrogen sulphides (H$_3$S) \cite{duan2014pressure} is crucial for understanding the origin of superconductivity in hydride materials. At pressures greater than 100 GPa and a critical temperature (T$_c$) of roughly 80 K, H$_3$S was predicted to be superconducting. Experiment later confirmed this theoretical prediction \cite{drozdov2015conventional}; H$_3$S is superconducting with a T$_c$ of 203 K at 200 GPa, which breaks the temperature record of cuprate of 164 K \cite{einaga2016crystal,mozaffari2019superconducting}. This was predicted by the Bardeen-Cooper-Schrieffer (BCS) phonon-mediated theory of superconductivity \cite{duan2014pressure,papaconstantopoulos2015cubic,errea2016quantum}. This implies that certain hydrogen molecules may exhibit superconductivity at extremely high temperatures. Finding new superconductors is still crucial in condensed matter physics. Expenses incurred in the development of materials can have superconducting at greater transition temperatures and perform in more realistic conditions.
\begin{figure*}
\begin{center}
  \includegraphics[width=0.90\textwidth]{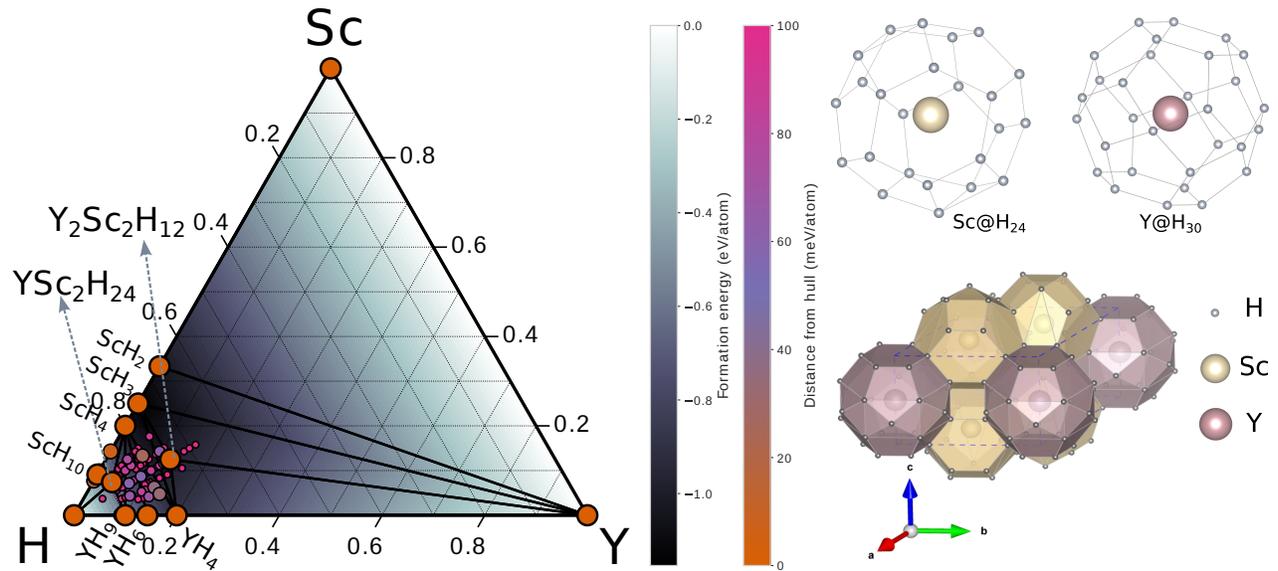}
\end{center}  
  \caption{(Color online) Convex hull representation for the Y–Sc–H system at 310 GPa. Solid black lines connect the thermodynamically stable phases. The colors within the regions of the convex hull indicate the overall formation enthalpy. The color of each circle reflects the distance of the associated phase from the convex hull Crystal structure of hexagonal YSc$_2$H$_{24}$ at 310 GPa, this structure contains Sc@H24 cage and Y@H30 cage. The Sc@H24 cage is composed of three hexagons and six pentagons, whereas the Y@H30 cage is composed of two hexagons, six rhombuses, and twelve pentagons.}\label{fig:crystal_structure} 
\end{figure*}
    Superconductivity at ambient conditions has long been difficult to achieve in the field of condensed matter science. Superconductivity was previously only observed at extremely low temperatures, often below 39 K which makes it impractical for liquid nitrogen cooling.\cite{kang2001mgb2,choi2002origin}. After discovering high-temperature superconductors \cite{bednorz1986possible,muller1987flux} such as cuprate superconductors with T$_c$ values as high as 164 K \cite{PhysRevB.50.4260}, scientists went out to find superconductivity at even higher temperatures. Whatever was done, the highest value of  T$_c$ remained constant for a long time. This demonstrates how difficult it is to make superconductivity work at room temperature. One method for searching for high-temperature superconductivity is to use hydride molecules with clathrate-like structures. Essential parts of these materials are elements located in the middle of the H$_{24}$ cages \cite{troyan2021anomalous,kong2021superconductivity,song2022systematic,song2021high}. They are known as core elements, and assist electrons pair up by acting as electron sources. Hydrogen atoms within these rings are held together by weak covalent interactions. In typical compounds like H$_3$S, strong covalent bonds connect hydrogen atoms to sulphur atoms. These new structures have undergone some modifications. YH$_{10}$ and rare-earth hydrides are predicted to have structures similar to clathrates but with additional hydrogen, namely H$_{32}$ cages\cite{heil2019superconductivity,semenok2021superconductivity}. Because they are so close to pure hydrogen, these superhydrides are expected to have high T$_c$. Theoretical calculation based on density functional theory predicted that the T$_c$ of YH$_6$ can reach 264 K at 120 GPa \cite{heil2019superconductivity,troyan2021anomalous} and 280 K at 200 GPa for LaH$_{10}$ \cite{kruglov2020superconductivity}. Recent work has explored various polyhydrides under high pressure, including studies by Chen et al. (2021) on lead hydrides \cite{chen2021phase}, Dou et al. \cite{dou2021ternary} on ternary Mg-Nb-H polyhydrides \cite{sun2023ternary}, Sun et al. on Na-PH superconductors\cite{sun2023ternary}, and Li et al. on alkaline boron hydrides \cite{li2022superconductivity}. These studies provide additional context for understanding the behavior of polyhydrides and underscore the relevance of high-pressure hydride research. Despite continuous efforts by scientists to discover materials exhibiting high-temperature superconductivity, conducting experiments that definitively validate the theoretical predictions remains challenging. Creating and studying superconducting phases under ideal conditions is a tough task that requires a considerable amount of resources and technological challenges. Thus, high T$_c$ research for metal hydrides has been particularly active in the recent decade. In this work, we are probing of potential stable alternative systems using ab-initio random structure searching with a fixed composition. We discovered a unique system with an H-cage structure that has significant electron phonon coupling, resulting in an anomalously high T$_c$ that may be used as a viable contender for room temperature superconductors.
\section{Computational methods}
The potential energy landscape of the Y-Sc-H system is investigated using Particle Swarm Optimization (PSO) as implemented in CALYPSO program \cite{wang2012calypso}. The PSO algorithm employs a swarm of 32 particles to locate thermodynamically stable structures, utilizing a PSO ratio of 0.6 to balance global and local search capabilities. The maximum number of iterations for the PSO is set to 100, which helps achieve optimal solutions while conserving computational resources. Our crystal structure search is limited to Y$_x$ = 1-3, Sc$_y$ = 1-3, and H ranging from 6 to 30, with a maximum total of 30 atoms. To construct the ternary convex hull, we used binary compounds: Y-H (YH$_4$, YH$_6$, YH$9$, YH${10}$) \cite{liu2017potential,peng2017hydrogen}, Sc-H (ScH$_2$, ScH$_3$, ScH$4$, ScH${10}$)\cite{peng2017hydrogen}, and the elemental phases $C2/c$-H\cite{pickard2007structure}, P6$_1$22-Sc\cite{akahama2005new}, and $Fddd$-Y\cite{buhot2020experimental}. The analysis was performed using Matador\cite{evans2020matador} and Pymatgen.\cite{ong2013python}
With found thermodynamically stable stoichiometric, we conducted further the ab-initio random structure searching (AIRSS)\cite{pickard2011ab} where we generated 2000 random atomic structures, ensuring that the minimum distance between atoms was at least 0.9 Å and did not exceed 3.2 Å. We employed AIRSS in conjunction with the Quantum ESPRESSO package \citep{QUANTUMESPRESO03_giannozzi2020quantum,QUANTUMESPRESSO02_giannozzi2017advanced} for performing geometry optimization based on Density Functional Theory (DFT) \citep{DFT01_hohenberg1964inhomogeneous,dft1965_kohn1965self}. The most stable structures resulting from AIRSS were then subjected to rigorous geometry optimization within Quantum ESPRESSO and electronic structure study. We utilized the generalized gradient approximation \cite{perdew1986accurate} in the form of the Perdew-Burke-Ernzerhof (PBE) \cite{PBE_perdew1996generalized} to determine the most energetically favorable atomic positions. To assess structural stability and the superconducting transition temperature, we conducted ab initio phonon calculations. These calculations involved the application of the McMillan formula, the Allen-Dynes formula, and the numerical solution of the Eliashberg function, which will be discussed in more detail in a later session. We carefully defined several computational parameters, including the cutoff energy for the plane-wave basis set and the grid mesh sizes over the Brillouin zone. Proper selection of these parameters is crucial for obtaining reliable results. Specifically, we set the energy cutoff for the plane-wave basis set and the charge density cutoff to 60 Ry and 480 Ry, respectively, to ensure convergence to within 1 meV/atom. We used a dense 18$\times$18$\times$18 k-mesh for self-consistent field calculations to achieve accurate phonon frequencies within a few $cm^{-1}$. Details of the phonon frequency convergence can be found in the Supplementary Information, Fig. S3. To enhance convergence, we applied the Marzari-Vanderbilt smearing scheme \cite{marzari1999thermal} with a broadening width of 0.02 Ry. For Density Functional Perturbation Theory (DFPT) \cite{baroni2001phonons} calculations, we utilized a 6×6×6 q-mesh to sample the reciprocal space and a much denser 36×36×36 k-point mesh for accurate electron-phonon coupling coefficients \cite{wierzbowska2005origins}. We employed pseudopotentials from the standard solid-state pseudopotentials (SSSP) library \cite{SSSP02_lejaeghere2016reproducibility} for the Y, Sc, and H elements.
\section{Main results and discussion}
We analyzed the decomposition enthalpies of the binary hydrides (Sc-H and Y-H) using a convex hull to establish the thermodynamic stability of Y-Sc-H, as shown in Fig. 1. The ternary convex hull reveals the
existance of 2 thermodynamically stable compounds which are YSc$_2$H$_{24}$ and Y$_2$Sc$_2$H$_{12}$.
Using these found stoichiometric, we conduct further random structure searching with AIRSS and add restrictions to ensure a minimum H-H distance of 1.0 \AA\ and a largest minimum distance of 3.2 \AA\ inside the system.
These constraints were particularly crucial for accommodating small distances, such as those associated with Hydrogen atoms.
The composition YSc$_2$H$_{24}$ inherits the MH$_6$, MH$_{10}$ (M=Y,La,Th,Sc) \cite{semenok2020superconductivity,du2022room,zurek2019high} and share the similar crystal structure with recent found LaSc$_2$H$_{24}$ superconductor \cite{he2024predicted}. Fig.S1 in the Supplementary Information shows the enthalpies of the 50 structures with the lowest values, revealing distinct plateaus in the enthalpy that strongly suggest the presence of structurally similar configurations, possibly corresponding to the same space groups. It should be acknowledged that the high number of atoms in the unit cell limited the success rate. The range of measured enthalpy was found to be vast, mostly due to the significant number of atoms present inside the system, as depicted in the inset of Fig. S1. Nevertheless, the utilisation of this structure search methodology has provided significant insights into the presence of different structures, particularly when considering the distinct enthalpy plateaus that are observed. We also evolutionary algorithm for crystal structure prediction \cite{yamashita2022hybrid} to ensure the accracy of the searching, the results are shown in Fig. S2 of the Supplementary Information.

\begin{figure}
\begin{center} 
\includegraphics[width=0.48\textwidth]{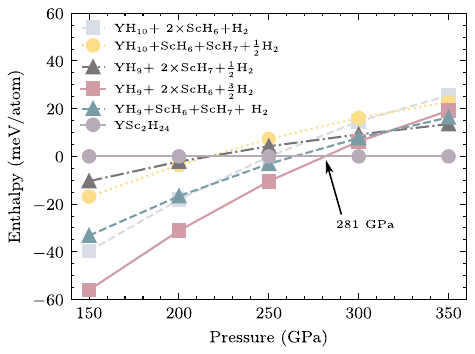}
\end{center}
  \caption{(Color online) A comparison of the enthalpies of YSc$_2$H$_{24}$ and several potential decomposition pathways originating from established structures, including Y$_{10}$, YH$_9$, ScH$_6$, ScH$_7$, and H$_2$, with respect to pressure. The arrow indicates the pressure where YSc$_2$H$_{24}$ becomes energetically favorable.}\label{fig:enthalpy_difference}
\end{figure}
Among these structures, the one with the lowest enthalpy was found as hexagonal YSc$_2$H$_{24}$ with the $P6/mmm$ space group. The optimized lattice constants at 310 GPa  are: $a$ = 4.60 \AA, $c$ = 3.24 \AA. Yttrium (Y) atoms are located in the centre of a Y@H30 cage in this configuration corresponding to Wyckoff site \textit{1b}, and the Y-H bonds have lengths of 1.89, 1.94, and 1.88, corresponding to three different types of H atoms. This cage structure is novel; it contains two hexagons, six rhombuses, and twelve pentagons (see Fig. \ref{fig:crystal_structure}). Sc@H24 is an additional cage-like structure composed of Scandium (Sc) atoms arranged in a 24-coordinate pattern, with Sc located at the corresponding Wyckoff site \textit{2c}; the Sc-H bonds in this configuration measure 1.80, 1.83 and 1.72 \AA, respectively. This Sc@H24 cage resembles the La@H24 and Y@H24 structures found in hydrogen cage \cite{song2021high,troyan2021anomalous,semenok2020superconductivity}. The crystal structure contains three distinct H sites positioned at the Wyckoff locations of $6m$ (0.76305 0.5261 0.5), $6j$ (0.22898, 0.0, 0.0), and $12n$ (0.37974, 0.0, 0.29064). The hydrogen-hydrogen bond lengths exhibit a range spanning from 1.06 to 1.17 \AA. Significantly, the H-H bond lengths in question display similarities to those observed in CaH$_6$ (1.24, 1.20, and 1.17 \AA\ at 150 GPa) \cite{jeon2022electron} and LaH$_{10}$ (1.07 \AA\ and 1.16 \AA\ at 300 GPa), hence leading to the establishment of an atomic hydrogen sublattice that has a strong resemblance to those present in other clathrate hydride structures. It is worth noting that these H-H distances are more extended than that of H-H in molecular hydrogen (0.74 \AA) \cite{grabowski2020molecular}. A detailed information of crystal structure of YSc$_2$H$_{24}$ at 310 GPa can be find in the Supplementary Information. We also compared the enthalpy of YSc$_2$H$_{24}$ with known structures, which can be a possible route to form this crystal at a range of pressure from 150-350 GPa as shown in Fig. \ref{fig:enthalpy_difference}. At pressures lower than 281 GPa, YSc$_2$H$_{24}$ is unlikely to form because of its higher enthalpy compared to a combination of YH$_{10}$ \cite{peng2017hydrogen}, ScH$_6$ \cite{ye2018high}, ScH$_7$ \cite{ye2018high}, and H$_2$ \cite{ozerov1962crystalline}. From 281 GPa and upwards, YSc$_2$H$_{24}$ has a lower enthalpy than these combinations, indicating its thermodynamic stability. It is worth noting, however, that the structure becomes dynamically stable only at pressures higher than 310 GPa within the harmonic approximation, as detailed in a later section. The phases used in enthalpy comparison are as follows, H$_2$: $C2/2$ (150-250 GPa), $Cmcm$ (300-350 GPa)\cite{pickard2007structure}, ScH$_6$: $P63/mmc$ (150-300 GPa), $Im\bar{3}m$ 350 GPa \cite{ye2018high}, ScH$_7$: $Cmcm$ \cite{ye2018high}, YH$_9$: $P63/mmc$ \cite{sun2019route}, and YH$_{10}$: $Fm\bar{3}m$ \cite{sun2019route}. Recent experiments demonstrate that YH$_{10}$ was not detected \cite{kong2021superconductivity}; however, with an enthalpy approximately 13-21 meV/atom lower than that of YH$_{10}$, YH$_9$, ScH$_6$, ScH$_7$, and H$_2$ combinations, YSc$_2$H$_{24}$ shows potential for realization in future experiments. This compound may be stabilized using high-pressure techniques, such as provided by a diamond anvil cell, starting perhaps from stable binary hydrides such as ScH$_{10}$ and YH$_6$. The YSc$_2$H$_{24}$ phase may be reproduced, too, by hydrogenation under controlled pressure and temperature conditions. We have also identified a new energetically stable structure with the stoichiometry Y$_2$Sc$_2$H$_{12}$. In this monoclinic structure, classified under the C2/m space group, hydrogen atoms form a complex network around the metal centers, resulting in various coordination geometries and bond lengths. Both Yttrium (Y) and Scandium (Sc) are encased in a shell of 15 hydrogen atoms, creating Y@H15 and Sc@H15 coordination environments, respectively. The crystal structure of Y$_2$Sc$_2$H$_{12}$ in the C2/m space group is depicted in Fig. S4 of the supplementary information, with detailed structural data provided in Table II of the supplementary information.
\begin{figure}
\begin{center}
  \includegraphics[width=0.48\textwidth]{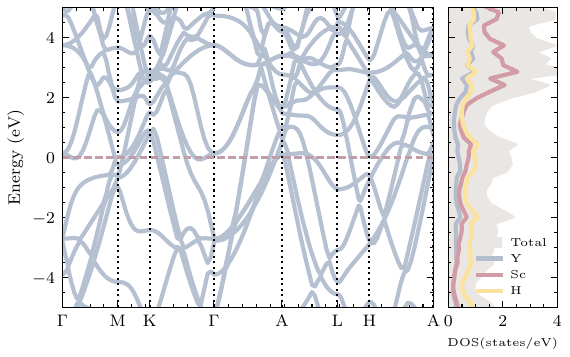}
\end{center}
  \caption{(Color online) Electronic band structure of YSc$_2$H$_{24}$ at 310 GPa and projected electronic density of states (PDOS) of each element. The dash line indicates the Fermi level.}\label{fig:band_structure}
\end{figure}

\begin{figure}
\begin{center}
\includegraphics[width=0.5\textwidth]{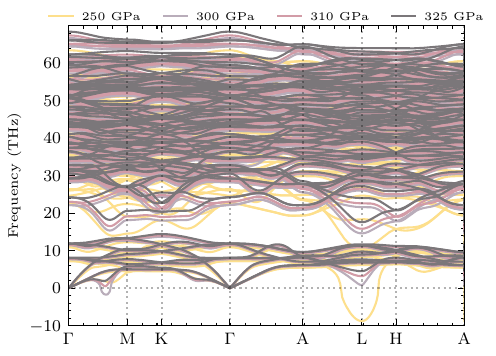}
\end{center}
  \caption{(Color online) YSc$2$H${24}$ phonon dispersion at different pressures ranging from 250 to 325 GPa. The region of imaginary frequencies marked by the zero-line frequency reflects the system's dynamic stability. A phonon mode experiences instabilities at 250 GPa, as evidenced by the imaginary mode at the L point of the Brillouin zone. While 300 GPa still shows signs of instability, with imaginary modes present at $\Gamma-M$ or a particularly sensitive imaginary mode at L, at 310 GPa and above, all frequency modes are real, indicating dynamic stability.}\label{fig:phonon_pressure}
\end{figure}

In order to investigate the electronic characteristics of YSc$_2$H$_{24}$ at a pressure of 310 GPa, we conducted calculations to determine its electronic band structure and projected electronic density of states (PDOS). The metallic nature of YSc$_2$H$_{24}$ is indicated by the presence of multiple electronic bands intersecting the Fermi level, as depicted in Fig. \ref{fig:band_structure}. The partial density of states (PDOS) analysis of YSc$_2$H$_{24}$ reveals a significant density of states (DOS) originating from H at the Fermi level. This resemblance to the DOS observed in LaH$_{10}$ \cite{verma2021prediction,quan2019compressed} and YH$_{10}$ \cite{heil2019superconductivity}, CaH$_6$ \cite {jeon2022electron} at a pressure of 310 GPa suggests that YSc$_2$H$_{24}$ will likely display high-temperature superconductivity primarily driven by hydrogen. The density of states (DOS) obtained from Sc has a similar pattern to the DOS derived from H, highlighting the significance of Sc's contribution to the superconductivity observed in this particular system. The partial density of states (PDOS) analysis shows that hydrogen atoms contribute significantly to the DOS at the Fermi level, accounting for 46.5\%. Sc atoms also contribute notably, with 35.6\%, while the contribution from Y atoms is less pronounced at 17.9\%. The significant presence of Sc’s DOS at the Fermi level indicates that Sc atoms may play an active role in enhancing electron-phonon coupling and, therefore, could positively impact the superconducting properties.
In metal-hydrogen cage structures, there is typically a charge transfer from the metal atoms to the H-atoms. This process not only serves as a stabilizing force for the cage structure but also supplies additional electrons for superconducting pairing. To quantify the amount of charge transfer, we utilize Bader charge analysis, as detailed in \cite{henkelman2006fast}. Our analysis reveals that Y donates 1.35e- to H-atoms, while each Sc atom donates 1.1 e- to H-atoms. On average, each H atom accepts 0.15 electrons, which proves insufficient for dissociating H$_2$ but significantly contributes to bond weakening. This effect results in a longer H-H distance, with the minimum distance between H-H found to be 0.8 \AA in this system. Consequently, YSc$_2$H$_{24}$ does not contain monatomic H atoms.

Due to their lightweight nature, hydrogen vibrations in the lattice are perhaps the most important driving source for the superconductivity of solid hydrogen and hydride superconductors. Fig. \ref{fig:phonon_pressure} shows the phonon dispersion of YSc$_2$H$_{24}$ under different pressures from 250 to 325 GPa. The space group of the structure remains the same within this pressure range, leading to quite similar dispersion curves. However, structural instability exists at the lower pressure of 250 GPa, located at the L-point of the Brillouin Zone. There is also a Kohn anomaly at q-point in between the direction $\Gamma-M$, which may indicate strong electron-phonon coupling or the possibility of instability. These characteristics vanish at higher pressures, such as 310 and 325 GPa. At 300 GPa, the L-point mode becomes real while an edge of instability is still present at q-point in between the direction $\Gamma-M$, and at 310 GPa and above, this behavior disappears completely. It is worth noting that the phonon dispersions of this compound are quite sensitive to k-point sampling. In this work, a 18×18×18 k-point mesh was found to converge the low-lying modes at the L points and the q-points near M, as shown in Fig. S3 of the Supplementary Information. At 250 GPa, the high frequencies ($>$ 500 cm$^{-1}$) in YSc$_2$H$_{24}$ approach and intersect with the lower-frequency range. However, at higher pressures, these high-frequency ranges form a phononic gap that separates them from the lower frequencies. As expected, an increase in pressure leads to the expansion of the frequency width, as clearly evident in Fig. \ref{fig:phonon_pressure}.

Recent theoretical works about Y-Sc-H compounds have predicted several stable phases with significant superconducting potential. For example, cP8-ScYH$_6$ has been proved dynamically stable even at pressure as low as 0.01 GPa, and a T$_c$ as high as 32.1 K means that good hydride superconductors can be obtained even under moderate pressure conditions \cite{wei2021formation}. On the other hand, Sc$_{0.5}$Y$_{0.5}$H$_6$, with a T$_c$ as high as 127 K, has shown that substitution of elements probably can further enhance superconductivity \cite{sukmas2022roles}. Further, the more hydrogen-rich phases, such as YScH$_{12}$ \cite{shi2024prediction}, have even higher critical temperatures, up to 179 K, which indicates the role of hydrogen in electron-phonon coupling strengthening. This is an important factor in superconducting behavior for YSc$_2$H$_{24}$s as well. \color{black}Thus, it is interesting to explore the electron-phonon interaction of this system, due to the existence of H-cage and its relation to the known high T$_c$ superconductor like YH$_{10}$ YH$_6$, and other Y-Sc-H compounds. Following we only discuss in detail the case 310 GPa. The electron-phonon coefficients can be obtained as \cite{baroni2001phonons}
\begin{equation}
g_{q\nu}(k,i,j) = \left(\frac{\hbar}{2M\omega_{q\nu}}\right)^{1/2}
\langle\psi_{i,k}| \frac{dV_{SCF}}{d {\hat u}_{q\nu}} \cdot
                   \hat{\epsilon}_{q\nu}|\psi_{j,k+q}\rangle.
\end{equation}
where $M$ denotes the atomic mass, ${\bf q}$ and ${\bf k}$ represent wave vectors, and $i$, $j$, and $\nu$ denote phonon modes and electronic energy band indices, respectively. The phonon linewidth $\gamma_{q\nu}$ can be obtained from $g_{q\nu}$ using:

\begin{equation}
\gamma_{q\nu} = 2\pi\omega_{q\nu} \sum_{ij}
                \int \frac{d^3k}{\Omega_{BZ}}  |g_{q\nu}(k,i,j)|^2 
                \delta(e_{q,i} - e_F)  \delta(e_{k+q,j} - e_F),
\end{equation}
The electron-phonon coupling (EPC) constant, $\lambda_{q\nu}$, for mode $\nu$ at wavevector $\mathbf{q}$ is defined as
\begin{equation}
\lambda_{q\nu} = \frac{\gamma_{q\nu}}{\pi \hbar N(e_F) \omega_{q\nu}^2}.
\end{equation}
\begin{figure}
\begin{center}
  \includegraphics[width=0.5\textwidth]{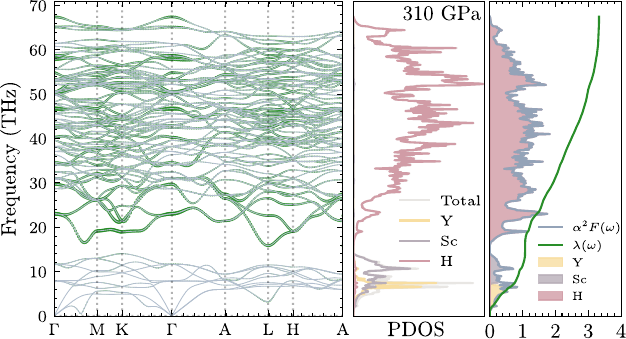}
\end{center}  
  \caption{(Color online) Phonon dispersion of YSc$_2$H$_{24}$, decorated with phonon linewidths which correspond to electron-phonon coupling strength, projected phonon densities of states, and Eliashberg spectral function at 310 GPa.}\label{fig:electron_phonon}
\end{figure}
We can also obtain the Eliashberg spectral function as $\alpha^2F(\omega) = {1\over 2\pi N(e_F)}\sum_{\mathbf{q}\nu}
\delta(\omega-\omega_{\mathbf{q}\nu})
{\gamma_{\mathbf{q}\nu}\over\hbar\omega_{\mathbf{q}\nu}}$, from which we can determine the total EPC by summing $\lambda = \sum_{\mathbf{q}\nu} \lambda_{\mathbf{q}\nu} =
2 \int {\alpha^2F(\omega) \over \omega} d\omega$. These pieces of information are crucial for estimating the superconducting transition temperature. However, before diving into this, let's explore these details further to enhance our understanding of this system. The similarity between the phonon density of states (DOS) and the Eliashberg spectral function underscores the phonon-mediated nature of superconductivity in this material, Notably, Y and Sc contribute significantly to the total EPC, especially at lower frequencies, with their contributions reaching up to one-third of the total EPC. Despite hydrogen atoms being the primary EPC contributors (67\%), the substantial impact of Y and Sc at lower frequencies, contributing to a $\lambda$ value up to 1.07, emphasizes the importance of heavy elements in enhancing superconducting properties. The calculated $\lambda$ value for YSc$_2$H$_{24}$ is 3.27 at 310 GPa which is larger than that of YH$_{10}$ (2.56) \cite{liu2017potential} (2.41) \cite{peng2017hydrogen} and similar to YH$_6$ (2.93) \cite{peng2017hydrogen} \cite{liu2017potential}. Its value is also comparable to LaH$_x$, which exhibits a wide range of $\lambda$ values, ranging from 1.44 to 3.94 \cite{kruglov2020superconductivity}, or Li$_2$MgH$_{16}$\cite{sun2019route}, which possesses a value near 4 over a similar range of pressure. Under greater pressure, at 325 GPa $\lambda$ reduces to 3.0 together with extra high phonon frequencies, making it somewhat similar to EPC properties at 310 GPa, as shown in Fig. \ref{fig:Eliashberg_function_310_325}.
\begin{figure}
\begin{center}
  \includegraphics[width=0.44\textwidth]{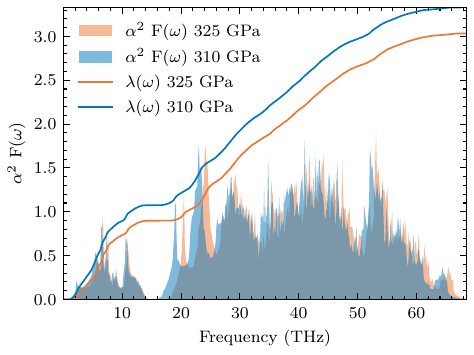}
\end{center}  
  \caption{(Color online) Eliashberg spectra function and cumulative $\lambda(\omega)$ at 310 and 325 GPa. The electron-phonon coupling strength of YSc$_2$H$_{24}$ decreases with increasing pressure.}.\label{fig:Eliashberg_function_310_325}
\end{figure} 
The calculated phonon linewidth, which is proportional to the Electron-Phonon Coupling (EPC), is depicted as circular markers on the phonon dispersion of YSc$2$H$_{24}$ in Fig. \ref{fig:electron_phonon}. At first glance, one can observe that most of the EPC results from the vibrations of hydrogen atoms, as evident in the Partial Density of States (PDOS, shown in Fig. \ref{fig:electron_phonon}), where the high-frequency range corresponds to hydrogen atom vibrations, reflecting their lightweight nature. In contrast, the heavier Yttrium (Y) and Scandium (Sc) atoms predominantly contribute to the lower frequency range.
With values of $\lambda$ greater than 1.5, which fall into the strong coupling regime, the McMillan formula is known to underestimate the $T_c$ \cite{allen1975transition}. Therefore, we revert to the Allen-Dynes equation to calculate $T_c$:
\begin{equation}
T_c = \frac{f_1f_2\langle\omega_{\text{log}}\rangle}{1.20} \exp\left(-\frac{1.04(1+\lambda)}{\lambda - \mu^\ast - 0.62\lambda\mu^\ast}\right)
\end{equation}
where $f_1$ and $f_2$ are the strong-coupling correction and the shape correction respectively \cite{allen1975transition}. 
\begin{align}
f_1 & = \left[1 + \left(\frac{\lambda}{2.46 (1 + 3.8\mu^{\ast})}\right)^{3/2}\right]^{1/3} \\
f_2 & = 1 + \frac{(\omega_2/\omega_{\text{log}} - 1)\lambda^2}{\lambda^2 + (1.82(1 + 6.3\mu^\ast)(\omega_2/\omega_{\text{log}}))^2}
\end{align}
The modified McMillan equation has exactly the same form but with $f_1f_2 = 1$. The obtained $T_c$ at 310 GPa is $T_c^{\text{McMillan}}$ = 194K (175K) with a Coulomb repulsive parameter $\mu^*$ of 0.1 (0.16) compared to $T_c^{AD}$ = 272K (229K) which clearly demonstrates the noted difference.
 We also used Matsubara-type linearized Eliashberg equations \cite{eliashberg1960interactions} to compute the critical temperature of the superconducting transition:
\begin{equation}
\hbar\omega_j = \pi(2j+1)k_BT, j=0,\pm1,\pm2,..
\end{equation}
\begin{equation}
\lambda(\omega_i-\omega_j) = 2\int^{\infty}_0\frac{\omega\alpha^2F(\omega)}{\omega^2+(\omega_i - \omega_j)^2}d\omega
\end{equation}
\begin{multline}
\Delta(\omega=\omega_i,T) = \Delta_i(T) = \\
\pi k_BT\sum_j \frac{[\lambda(\omega_i-\omega_j) - \mu^*]}{\rho+\left| \hbar\omega_j + \pi k_BT\sum_k (sign \omega_k). \lambda(\omega_j - \omega_k) \right|}.\Delta_j(T)
\end{multline}
\begin{table*}[!htb]
\centering
\caption{Characteristic superconducting properties of $P6/mmm$ YSc$_2$H$_{24}$ computed for different values of the Coulomb pseudopotential $\mu^*$ at pressures of 310 and 325 GPa.}
\label{tab:superconductivity}
\begin{tabular}{lrrrr}
\hline
\hline
Parameter & 310 GPa ($\mu^* = 0.1$) & 310 GPa ($\mu^* = 0.16$) & 325 GPa ($\mu^* = 0.1$) & 325 GPa ($\mu^* = 0.16$) \\
\hline
McMillan equation $T_c^{McMillan}$ (K) & 177.0 & 160.5 & 187.6 & 169.1 \\
Allen-Dynes equation $T_c^{AD}$ (K) & 263.9 & 221.3 & 266.5 & 224.0 \\
Eliashberg equation $T_c^{Eliashberg}$ (K) & 330 & 302 & 309 & 293 \\
$\lambda$ & 3.27 & 3.27 & 3.00 & 3.00 \\
$\omega_\text{log}$ (K) & 948.8 & 948.8 & 1043.0 & 1043.0 \\
Sommerfeld constant (mJ/mol$\cdot$K$^2$) & 11.7 & 11.7 & 11.0 & 10.9\\
Superconducting Gap (meV) & 77.7 & 72.2 & 74.0 & 70.0\\
Characteristic ratio (2$\Delta$/k$T_c$) & 5.5 & 5.5 & 5.5 & 5.5\\
Specific heat jump/$T_c$ (mJ/mol$\cdot$K$^2$) & 12.2 & 20.9 & 24.3 & 26.9\\
Upper critical field (T) & 84.5 & 32.1 & 84.4 & 81.9\\
\hline
\end{tabular}
\end{table*}
Where T is the temperature (K), $\omega$ is the frequency (Hz), $\rho(T)$ is a pair-breaking parameter, and the function $\lambda(\omega_i-\omega_j)$ corresponds to effective electron-electron interaction via phonon exchange \citep{bergmann1973sensitivity}. The superconducting transition temperature can be calculated as a solution of the equation $\rho(T_c)$ = 0, where $\rho(T)$ is defined as $max(\rho)$ if $\Delta(\omega)$ is not a zero function of $\omega$ at a fixed temperature. The above equations can be written in a matrix form \cite{allen1983theory,kruglov2020superconductivity}:
\begin{multline}
\rho(T)\psi_m=\sum_{n=0}^N K_{mn}\psi_n\Leftrightarrow\rho(T)\begin{pmatrix}
\psi_1\\
...\\
\psi_N
\end{pmatrix}=\\
\begin{pmatrix}
K_{11}& ... & K_{1N}\\
...& K_{ii} & ...\\
K_{N1}& ... & K_{NN}
\end{pmatrix} \times \begin{pmatrix}
\psi_1\\
...\\
\psi_N
\end{pmatrix}
\end{multline}
where $\psi_n$ relates to $\Delta(\omega,T)$ and K$_{mn}$ is defined as:
\begin{multline}
K_{mn} = F(m-n)+F(m+n+1)=\\
2\mu^*-\delta_{mn}\left[ 2m+1+F(0)+2\sum_{l=1}^mF(l)\right]
\end{multline}
with
\begin{equation}
F(x)=F(x,T) = 2\int_0^{\omega_{max}}\frac{\alpha^2F(\omega)}{(\hbar\omega^2)+(2\pi.kT.x)}\hbar^2\omega d\omega
\end{equation}
where $\delta_{nn}=1$ and $\delta_{mn} = 0(n\neq m)$ is a unit matrix. The vanishing of the maximum eigenvalue of the matrix $K_{nm}$ can now be used to replace the criteria of $\rho(T_c)$ = 0 to find T$_c$. The present approach prevents the necessity of directly solving the system of nonlinear integral equations, hence eliminating concerns over the selection of an initial approximation to the solution and the convergence of the algorithm. One of the drawbacks associated with this particular approach is the potential uncertainty, often within a range of $\pm$5 K. Additionally, it is not feasible to determine the magnitude of the superconducting gap at 0 K using this method. Therefore, this approach is appropriate for acquiring a reliable estimate of $T_c$ for subsequent computations using more computationally intensive techniques \cite{margine2013anisotropic}. Nevertheless, we applied this simple method to other known systems like YH$_6$, LaH$_{10}$, YH$_{10}$, H$_3$S, the results are comparable to previous studies. Furthermore, the Sommerfeld constant can also be obtained from:
\begin{equation}
\gamma =  \frac{2}{3}\pi^2k_B^2N(0)(1+\lambda)
\end{equation}
which was utilised to calculate the upper critical magnetic field $H_c(0)$ and the specific heat jump $\Delta C$ \cite{carbotte1990properties}:
\begin{equation}
\frac{\gamma T_c^2}{H_c^2(0)}=0.168\left[ 1-12.2\left( \frac{T_c}{\omega_{log}}\right)^2ln\left (\frac{\omega_{log}}{3T_c}\right)\right]
\end{equation}
\begin{equation}
\frac{\Delta C(T_c)}{\gamma T_c} = 1.43\left[ 1+53\left( \frac{T_c}{\omega_{log}}\right)^2 ln\left( \frac{\omega_{log}}{3T_c}\right)\right]
\end{equation}
The superconducting gap can also be be estimated using a semi-empirical equation \cite{carbotte1990properties}, as long as it satisfied the condition $T_c/\omega_{log} < 0.25$:
\begin{equation}
\frac{2\Delta(0)}{k_BT_c} = 3.53\left[ 1+12.5\left( \frac{T_c}{\omega_{log}}\right) ln\left( \frac{\omega_{log}}{2T_c}\right) \right]
\end{equation}
With the obtained superconducting gap, we can see the change as a function of temperature as shown in Fig. \ref{fig:Superconducting_gap} using the phenomenological formula:
\begin{equation}
\Delta(T) = \Delta(T_0)\sqrt{1-\left( \frac{T}{T_c}\right)^k}
\end{equation}
where $k$ =3 in the framework of BCS theory \cite{eschrig2001theory}.
\begin{figure}
  \includegraphics[width=0.45\textwidth]{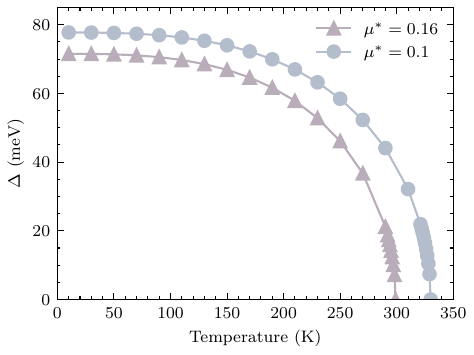}
  \caption{(Color online) The superconducting gap of YSc$_2$H$_{24}$ as a function of temperature for two values of $\mu^*$, at 310 GPa, the superconducting transition temperature locates where the gap becomes zero.}.\label{fig:Superconducting_gap}
\end{figure}
As can be seen from Fig. \ref{fig:Superconducting_gap}, within the acceptable range of $\mu^*$ from 0.1 to 0.16, the superconducting gap vanishes at temperatures exceeding room temperature within the solution of the Eliashberg equation, which is very encouraging. At higher pressures, $T_c$ continues to maintain high values, as in the case of 310 GPa. A summary of calculated superconducting properties of YSc$_2$H$_{24}$ is shown in Tab. \ref{tab:superconductivity}. The $C2/m$ phase of Y$_2$Sc$_2$H$_{12}$ is dynamically stable, as evidenced by its phonon dispersion, shown in Fig. S5. Similar to the cage-like structure observed in YSc$_2$H$_{24}$, this phase features Y@H15 and Sc@H15 coordination environments. However, its band structure reveals a significantly lower density of states (DOS) at the Fermi level, measured at 6.9 states/spin/Ry/unit cell compared to 15.8 for YSc$_2$H$_{24}$. Additionally, the electron-phonon coupling (EPC) is relatively weak, as detailed in Fig. S6 of the supplementary information, contributing to a much lower superconducting transition temperature T$_c$ of 7.4 K.
\section{Conclusion}
To summarize, we employed ab-initio random structure searches (AIRSS) in conjunction with Particle Swarm Optimization (PSO) to discover the new hexagonal and thermodynamically stable phase YSc$_2$H$_{24}$ at high pressure. The new structure has a one-of-a-kind hydrogen clathrate network made up of peculiar Sc@H24 and Y@H30 cages. Intriguingly, phonon dispersion and electron-phonon coupling calculations suggest a superconducting T$_c$ of up to 330 K at 310 GPa, which is higher than ambient temperature. Our finding expands the diversity of clathrate hydride structures and enriches the prospects for near-higher room temperature superconductors, which can help increase future theoretical and experimental research in the field. 
\section*{Conflicts of interest}
The authors declare that they have no conflict of interest.
\section*{Data availability}
The data supporting this article have been included as part of the Supplementary Information.
\section*{Acknowledgements}
We would like to express our appreciation to Prof. Ching-Ming Wei at the Institute of Atomic and Molecular Sciences at Academia Sinica in Taipei, Taiwan for generously providing us with the computational resources needed to complete this study.



\balance


\providecommand*{\mcitethebibliography}{\thebibliography}
\csname @ifundefined\endcsname{endmcitethebibliography}
{\let\endmcitethebibliography\endthebibliography}{}

\bibliographystyle{rsc} 

\end{document}